\documentstyle[prd,aps,preprint]{revtex}
\begin{document}
\draft
\title{ Quenched QCD at finite density}

\author{J.~B.~Kogut , M.-P.~Lombardo}
\address {Physics Department, University of Illinois at Urbana-Champaign,
Urbana, IL 61801-3080}
\author{D.~K.~Sinclair}
\address{HEP Division, Argonne National Laboratory, 9700 South Cass Avenue,
Argonne, IL 60439}
\date{\today}
\maketitle

\begin{abstract}

Simulations of quenched $QCD$ at relatively small but
{\it nonzero} chemical potential $\mu$ on
$32 \times 16^3$ lattices
indicate that the nucleon
screening
mass decreases linearly as
$\mu$ increases predicting a critical
chemical potential of one third the
nucleon mass, $m_N/3$, by extrapolation.
The meson spectrum does not change as $\mu$ increases
over the same range, from zero to $m_\pi/2$.
Past studies of quenched lattice QCD have suggested
that there is phase transition
at $\mu = m_\pi/2$. We provide alternative explanations
for these results, and
find a number of technical reasons why standard lattice simulation
techniques suffer from greatly enhanced fluctuations
and finite size effects for $\mu$
ranging from $m_\pi/2$ to $m_N/3$. We find evidence for such
problems in our simulations, and suggest that they can
be surmounted by improved measurement techniques.
\end{abstract}
\newpage
\setcounter{page}{1}
\pagenumbering{arabic}
\section{Introductory Comments}
{}~~~~The search for the quark gluon
plasma (QGP) is one of the major challenges
posed by QCD, both theoretically and experimentally
\cite{QGP}. At the experimental
level, no compelling evidence of this new state of matter has yet been found.
Results are expected, however, from
the experiments at RHIC. Most of the underlying physics of the
QGP can be studied theoretically and computationally
only by lattice simulations \cite{CHEM1} \cite{CHEM2}
 at this time. These have been quite successful
in describing the physics at zero density and finite temperature, while
simulations at finite density and zero temperature have been plagued in the
past by many uncertainties \cite{BAR}.

First, simulations of QCD with virtual quarks
are difficult because the action is complex at nonzero $\mu$,
thus preventing
the na\"{\i}ve use of probabilistic methods in the evaluation of functional
integrals. Nonetheless, complex Langevin simulations of spin models which
are related to the strong coupling limit of lattice QCD have been
quite encouraging  \cite{CL}.
 We will see below that complex Langevin simulations
of lattice QCD will, however, face numerical and conceptual problems which
are not contained in toy spin models.

Second, while quark models of nuclear matter predict that the nucleon
screening
mass will decrease linearly with increasing chemical potential $\mu$ and a
chiral symmetry restoring transition will occur at $\mu_c=m_N/3$
where the nucleon becomes massless,
past simulations of the {\it quenched} theory
have suggested that,
in the limit of massless quarks,
the system is in the
deconfined, chirally symmetric phase no matter how small the chemical potential
$\mu$ is
\cite{BAR}! When massive quarks have been simulated, the results have
suggested that the critical chemical potential is $m_\pi/2$. This has
caused most workers in the field to claim that quenched QCD is unphysical
at nonzero $\mu$. However, as we will discuss more fully below,
there are several explanations for this apparently
pathological behaviour which are unrelated to quenching: for instance,
confinement is essential to obtain the correct $\mu$ dependence
in QCD, but it is a
property of the ensemble average of configurations, and
need not be apparent configuration by configuration;
the lattice spacing used in the simulations has been
too coarse, and, as a consequence, flavor symmetry
breaking caused  by staggered fermions may play an
important role and suppress one's estimate of $\mu_c$ \cite{EM1};
finite volume effects are very large at finite $\mu$ \cite{SU3_1}; and, quark
propagator algorithms are very slowly convergent at nonzero $\mu$
\cite{CHEM1}.

Attempts to clarify these issues by studying simpler models
have not been decisive. Although the quenched approximation
failed qualitatively for single site models based on the gauge
group U(1) \cite{U1}, it proved to be a good guide to such models based
on SU(3) \cite{SU3_1}.
A number of studies have shown that lattice artifacts
are particularly large and dangerous at nonzero $\mu$ \cite{SU3_1}
\cite{VINK}.
Analytic arguments for lattice QCD
have been proposed which suggest
that the correct behavior is recovered in the continuum \cite{SU3_2}.

Based on these considerations, we have decided to re-examine
the quenched theory in greater detail, and have successfully completed
a first round of simulations \cite{LAT93}.
As will be discussed in the text which follows,
our results, obtained for values
of chemical potentials ranging from zero to half the pion mass,
are consistent with a critical chemical potential of one third the
nucleon mass, expected on physical grounds.

Unfortunately, we have not been able to
adequately
explore the more interesting region
$\mu > m_\pi/2$ with our available resources.
However,we believe that by using larger lattices and
measurement techniques better tuned to the physics of nonzero
chemical potential, we will be able to simulate the model
successfully for chemical potentials closer to $m_N/3$.

This  paper is organized into three additional sections. In Sec.II
we discuss several reasons why the region of chemical potential
between $m_\pi/2$ and $m_N/3$ is difficult to simulate by traditional
lattice gauge theory methods which are successful at zero $\mu$. In
Sec.III we present our new $32 \times 16^3$ simulation data. These
results include the first spectroscopy calculations at nonzero $\mu$.
In Sec.IV we summarize our results and give our strategies and plans
for future work which, we hope, will allow us to simulate chemical
potentials closer to the critical point.

\section{Simulation Problems at and beyond $\mu = m_\pi/2$}
{}~~~~First recall the well studied case of zero chemical potential
(zero baryon number density). The normal QCD (lattice) action is quadratic
in the fermion fields, so one can perform the fermion integral explicitly
leading to the determinant of the Dirac operator

\begin{equation}
    M_{n,n'} = \frac{1}{2} \sum_\lambda \eta_{n \lambda}(U_{n,\lambda}
    \delta_{n+\lambda,n'}-U^\dagger_{n',\lambda} \delta_{n-\lambda,n'})
    + m_q \delta_{n,n'}
\end{equation}
where $n$ and $n'$ label nearest neighbor sites, $U$ are the
$SU(3)$ gauge fields on the links and
$\eta$ the staggered quark phase factors. The new gauge action
for $N_f$ flavors is then

\begin{equation}
S = \sum_{\Box} \beta (1-\frac{1}{3}
{\mathop{\rm Tr}}_{\Box} UUUU) - \frac{N_f}{4}
{\mathop{\rm Tr}}(\ln M)
\end{equation}
which produces a partition function

\begin{equation}
    Z = \int {\cal D}U e^{-S}
\end{equation}
The expectation value of any observable f(U) is given by
\begin{equation}
   \langle f \rangle = \frac{1}{Z}\int {\cal D}U f(U) e^{-S(U)}
\end{equation}

In the quenched approximation one sets $N_f$ to zero in the partition
function. Standard Monte Carlo methods then apply to the numerical
evaluation of expectation values and have been quite successful
at vanishing $\mu$.

     Now we turn to the case where there is a finite chemical potential $\mu$
for quark number. This is imposed by making the replacement
\begin{equation}
              U_{n,4} \rightarrow e^{\mu} U_{n,4}
\end{equation}
and
\begin{equation}
              U^\dagger_{n,4} \rightarrow e^{-\mu} U^\dagger_{n,4}
\end{equation}
in the definition of M. This adds the complication that
${\mathop{\rm Tr}}(\ln M)$ is no
longer real and the exponent in the definition of Z develops a phase
when $N_f$ is nonzero \cite{CHEM1}. This problem provides additional incentive
to pursue the quenched approximation since complex actions and their
attendant simulation methods, such as complex Langevin algorithms, are
not well understood.

    When $\mu$ is nonzero we see from the expression
for the Dirac operator that quark propagation in the positive
``time'' direction is favored. In a diagrammatic expansion
of an expectation value involving quarks there will be
closed loops of quarks winding preferentially in the positive
time direction. As discussed and illustrated in detail in \cite{CHEM1}, at
sufficiently large $\mu$ large quark loops winding all the way around
the periodic lattice in the time direction will appear.
Viewing a time-slice of the partition
function, this means that positive $\mu$ will favor a ground
state with a net baryon density.

     There are extra complications involved in inverting the
Dirac operator when $\mu\neq 0$. A row in the inverse of the Dirac operator
is needed in spectroscopy and chiral condensate calculations.
If we use the conjugate gradient algorithm to invert
$M$, we do this by inverting $M^{\dagger}M$ on the source multiplied by
$M^\dagger$. This is necessary, since the conjugate gradient algorithm
is designed for
positive definite matrices. From the definition of $M$ given above, it is
clear that for $\mu=0$, $M^{\dagger}M$ is block diagonal, connecting even sites
to even sites, and odd sites to odd sites. This halves the amount of work one
might na\"{\i}vely have expected to perform. No such symmetry exists for
$\mu\neq 0$.

     For the $\mu=0$ case the diagonal term in $M$ is hermitean, the hopping
term is skew-hermitean. Thus all eigenvalues have real part $m_q$. The minimum
eigenvalue of $M^{\dagger}M$ is $\geq m_q^2$ and convergence of the conjugate
gradient is guaranteed. For $\mu\neq 0$ no such simple analysis is possible.
Small eigenvalues are possible and $M^{\dagger}M$ is relatively
ill-conditioned. If no winding of the quark lines around the lattice in the
time direction is possible, the matrix elements will be simply related to their
$\mu=0$ counterparts, and the fermion determinant will remain real and
proportional to its $\mu=0$ value. Once winding occurs, the system acquires
a ground state with non-zero baryon number density and physics changes.
Although the conjugate gradient algorithm continues to work when $\mu$
is nonzero, it converges very slowly. For example, requiring
the ``residual'' on a $32 \times 16^3$ lattice be
less than $10^{-6}$ then causes the conjugate gradient routine to use
approximately 650 sweeps to converge when the coupling is $\beta = 6.0$,
the bare quark mass is $m_q = .02$, and the chemical potential vanishes.
When $\mu$ is increased to 0.10, approximately 1,500 sweeps are needed
for convergence. At $\mu = .15$ that number grows to 5,000, and at
$\mu = .17$ it is typically 8,500. Many
past studies of quenched QCD at nonzero $\mu$ have not faced up to the
slow convergence of iterative algorithms to invert $M$. In fact,
a number of Lanczos studies \cite{BAR1}
simply noted that the computer time needed
to find the physically relevant small eigenvalues of $M$ grew prohibitively
large at nonzero $\mu$, and only the largest eigenvalues were obtained
\cite {DavKle}. These partial results motivated us to study the
stability and convergence of the conjugate gradient algorithm
\cite {AIM} as a function of the stopping residual.
We are, therefore, confident that the results we present in Sec.II
below are as reliable as possible.

    Past studies of quenched QCD have also noted that problems begin
to appear in their calculations when the chemical potential approaches
$m_\pi/2$. We shall argue now that the two most important features
of QCD, chiral symmetry breaking with a Goldstone pion and confinement,
are responsible for these difficulties. We will see reasons why
traditional lattice gauge theory calculational methods become very
inefficient at nonzero $\mu$, and we will suggest minor ways to improve them.
One set of problems arises because the expected result $\mu_c = m_N/3$
relies on confinement which is a property of an ensemble average
rather than a property of single configurations on which we make
measurements. We will argue that there are spurious effects in the
quark propagators calculated on individual gauge configurations
which are large at substantial $\mu$ and yet should cancel in
ensemble averages by virtue of confinement. One telltale symptom of such
effects
is that the approximate realization of continuum symmetries on individual
configurations is no longer manifest.
A second set of problems
arises because the natural dispersion in $m_N/3$ and $m_\pi/2$
estimates calculated on individual configurations overlap for
lattice sizes typically used at present.

    To understand why one might expect algorithmic problems in
calculations of hadron propagators near $\mu = m_\pi/2$ when they
are calculated on individual configurations, it is simplest
to consider point-to-point hadron propagators for a fixed source
and sink on the lattice. For this discussion we will consider only the
exponential behavior and ignore the power-law multiplier. At $\mu = 0$
the average meson propagator has two terms at large separations $T$ and
$N_t-T$, one proportional to $exp(-m_\pi T)$, the other proportional to
$exp(-m_\pi(N_t-T))$.
(Here, and in what follows ``proportional to'' ($\propto$) is used to mean
proportional to up to a T dependent phase, or in the case of the quark Green
function, being a matrix in colour space, up to a T dependent matrix in colour
space of unit norm (the norm of a matrix $A$ is defined by
$\sqrt{{\mathop{\rm Tr}} A^{\dagger} A}$.))
Empirically the pion propagator measured on individual configurations is also
well approximated by 2 such terms, so we assume such an asymptotic form for
the point-point meson propagator on a typical configuration.
However, a meson propagator from point
$x= ({\bf x},0)$
to point $y=({\bf x},T)$ on a given configuration is just
$ {\mathop{\rm Tr}}(G(y,x)G(x,y)) \propto
{\mathop{\rm Tr}}(G^\dagger(x,y)G(x,y))$
where $G(x,y)$ is the quark propagator. This
means that the quark propagator on a typical configuration must also have 2
terms, one proportional to $exp(-\frac{m_\pi}{2} T)$, the other to
$exp(-\frac{m_\pi}{2}(N_t-T))$, corresponding to the quark propagating from
y to x in the 2 different time directions allowed by the periodic lattice.
We see immediately that this requires the meson
propagator on such a configuration to have a third term, whose magnitude is
less than or equal to $constant \times exp(-\frac{m_\pi}{2} N_t)$. (This
constant is the geometric mean of the magnitudes of those for the first and
second terms so that this statement has content.)
This term is the contribution where the quark and
antiquark go around the lattice in opposite directions annihilating when they
meet. Since such a term is {\it not} allowed by confinement, contributions from
different configurations must contribute with random phases and so cancel. Of
course, for $\mu=0$, this term is vanishingly small for large $N_t$ as are
terms coming from the quark winding multiple times around the lattice, so that
the meson propagator remains the sum of 2 terms.
The important point is that
confinement is {\it not} a property of a
single configuration (we consider the effect
of translating x about the lattice as considering multiple configurations
differing only by translation), but rather a result of the ensemble average.

     Now let us turn to the case where $\mu \neq 0$. Here the meson propagator
${\mathop{\rm Re}}
{\mathop{\rm Tr}}(G_\mu(y,x)G_\mu(x,y)) \propto
{\mathop{\rm Tr}}(G^\dagger_{-\mu}(x,y)G_\mu(x,y)$,
where the inclusion of the ${\mathop{\rm Re}}$
is the result of averaging over the given
configuration and that obtained by time reversal, which is equivalent to
taking $\mu \rightarrow -\mu$. For $\mu$ sufficiently small, one can use
the $\mu=0$ form for the quark propagator discussed above to argue
that the quark propagator $G_\mu$ will have 2 terms, one proportional to
$exp(-(\frac{m_\pi}{2}-\mu) T)$ and the other to
$exp(-(\frac{m_\pi}{2}+\mu)(N_t-T))$.
This means that while the first 2 terms
in the meson propagator will be as before, the third term will be replaced by
2 terms the more important of which has magnitude less than or equal to
$constant \times exp(-(\frac{m_\pi}{2}-\mu)N_t)$. Again, we expect such a term
to cancel between configurations because of confinement. However, as $\mu$
approaches $m_\pi/2$ this term is no longer small, and for $\mu > m_\pi/2$
it, in fact, becomes large! Thus we expect the behavior of the meson
propagator to change near
$\mu=m_\pi/2$, varying greatly from configuration to configuration.

Now,
quenched QCD has a global $Z_3$ symmetry which means (on the lattice) that
the 3 gauge configurations differing only by having the gauge fields pointing
in the +t direction from the top timeslice multiplied by a common element of
$Z_3$, occur with the same weights in the ensemble. It is easy to see that
averaging over these triplets of gauge configurations removes this third term,
and all terms where the quark line winds around the lattice, except where it
winds around the lattice a multiple of 3 times. The case where the quark line
winds exactly 3 times around the lattice describes the configuration where
the meson consists of a baryon-antibaryon pair which go round the lattice in
opposite directions. Such a state is allowed, but contributes a term
proportional to $exp(-(m_B-3\mu)N_t)$. On a single configuration we would have
predicted this state to behave as $exp(-3(\frac{m_\pi}{2}-\mu)N_t)$, which
again
becomes large near $\mu=m_\pi/2$. This ultralight 3-quark state on a typical
configuration, presumably representing a state of 3 unbound quarks,
must average to zero over the ensemble, as a consequence of confinement,
as must the contributions of any non-colour-singlet terms
leaving only the baryon contribution. Similar arguments can be applied to
6,9,... quark states.

      Similar arguments indicate that $< \bar{\psi}\psi >$ for individual
configurations will start acquiring extra contributions due to precocious
winding of quark lines around the lattice near $\mu = m_\pi/2$, but confinement
will require these to vanish in the ensemble average for $\mu < m_N/3$.
Similar effects will occur for the baryon propagator. The discussion here is
more complex, since there is an additional cancellation due to confinement
which occurs even at $\mu=0$.
The leading behaviour of the propagator for a
3-quark state would be expected to behave like $exp(-\frac{3 m_\pi}{2}T)$ on a
single configuration from our discussion above. Hence there must be
cancellations of this leading behaviour which describes the propagation of
3 free quarks, if we are to get the required $exp(-m_N T)$ behaviour.
Much of this cancellation occurs when we project the required colour singlet
state and average the sink over the timeslice to obtain
a zero momentum state. The rest of the cancellation must occur when the
ensemble average binds these 3 quarks into a baryon.

      Hence, even when the finite $\mu$ transition occurs at $m_N/3$ as
expected, one expects to find a great increase in the statistical fluctuations
of the hadron propagators near $\mu = m_\pi/2$. This is due to the fact that
full confinement is not realized on a configuration by configuration basis
but is rather a property of the ensemble average. Averaging over the three
$Z_3$
boundary conditions in the t direction is expected to reduce these fluctuations
by enforcing the $Z_3$ but not the $SU(3)$ requirements of confinement. Using,
as we do, not point-point, but rather point-zero momentum (or wall-zero
momentum) propagators could potentially give us some aspects of confinement.
As will be discussed further in Sec.III, we explicitly average over the
three $Z_3$ boundary conditions for each gauge configuration in an
attempt to enforce the $Z_3$ requirements of confinement.
However, our evidence is that this is insufficient to yield all aspects of
confinement on a single configuration.

    Another problem that can lead to considerable suppression of an
estimate of $\mu_c$ found in a low statistics calculation follows from
the known, large fluctuations seen in calculations of $m_N$. As will be
discussed in Sec.III, the distributions of $m_N/3$ and $m_\pi/2$ measured
on each configuration overlap even at $\mu = 0$. So, on some configurations
$m_N/3$ will be as small as $m_\pi/2$ measured on the same set of
configurations! In other words, the  spreading in $m_N$ will
suppress estimates of $\mu_c$ to the neighborhood of $m_\pi/2$ on $32 \times
16^3$ lattices at $\beta$ = 6.0.
This is a conventional
finite size rounding effect which should be lessened by simulating
larger lattices.

For individual configurations, quark lines can wind
multiple times around the lattice, even below the transition. Because the
contributions of such configurations can be very large they can dominate the
averages over a relatively small ensemble, giving false indications of
having entered the baryon rich phase.

Such winding is the source of large fluctuations. One characteristic of such
fluctuations is the fact that
the hadron propagators need no longer obey the symmetries of the
ensemble average (such as time reversal invariance) on a configuration by
configuration basis.

Finally, let us discuss how our scenario might appear in the approach of
\cite {BAR1} \cite{DavKle} which first pointed out several possible
pathologies in quenched QCD. The authors used the Lanczos algorithm
to determine several features of the eigenvalue distribution of $M$ on
a very small ensemble of $16^4$ configurations at $\beta = 6.2$\cite{DavKle}.
Much of their work concentrated on $\Delta$, the half-width of the eigenvalue
distribution of $M$.
 If we are correct the outer eigenvalues of $M(m_q=0)$ which they
calculate and base their criticism on
 would correspond to those modes for the Dirac equation on a single
configuration which cancel in observables when the ensemble average enforces
confinement.
 Our scenario
 would then require that at small $\mu$
 the eigenvalues of $M$ on the imaginary axis and, in
particular, near the origin  have a similar distribution to that at $\mu=0$.
These eigenvalues, which are the \underline {only} ones of direct
physical significance are difficult to calculate \cite{DavKle} and
little is know about them.

 Attempts have been made to
calculate the distribution of small eigenvalues of $M$ from those of
$M^{\dagger}M$. However, this method has potential problems. If we were indeed
considering zero, and thus degenerate eigenvalues of a matrix such as $M$,
$M^{\dagger}M$ could have a lower degeneracy of zero eigenvalues than $M$. The
reason is that, if $M$ has n zero eigenvalues, it will in general have only
$m\leq n$ eigenvectors with eigenvalue zero (i.e. an {\it incomplete} set of
eigenvectors), in which case $M^{\dagger}M$ has only m zero eigenvalues, not n.
When this degeneracy is broken so that $M$ has n small eigenvalues, and a
complete set of eigenvectors, $M^{\dagger}M$ will still have only m small
eigenvalues. The fact that $M^{\dagger}M$ might have more of its eigenvalues
far from the origin than $M$ should come as no surprise since $M^{\dagger}M$
admits contributions to observables where a quark-antiquark pair winding once
around the lattice together is enhanced by a factor of $exp(2\mu N_t)$.
These contributions are absent for $M$.

New Lanczos studies of the eigenvalues of $M$ would be very instructive,
especially if they were accompanied by conjugate gradient calculation
of $<\bar\psi\psi>$, $<J_0>$ and spectroscopy, configuration by configuration.
\section{The Simulation}

\subsection{Observables}
{}~~~~We first describe the measurements we have performed, with emphasis
on the special features of the theory at finite density.
They are interesting, since some give rise to relationships which
hold exactly configuration by configuration, and are useful to check the
convergence of the inversion. Others imply relationships which must be
true only in the infinite statistics limit, and are useful to check
the quality of our data sample. All of them follow
from the modified symmetries of the Dirac operator:
\begin{equation}
M^{\dagger}_\mu = -M_{-\mu}
\end{equation}
\noindent
or, equivalently, from the transformation of the quark propagator
$G_\mu$ under time reversal:
\begin{equation}
T(G_\mu(0;n)) = G_\mu(n;0) = (-1)^nG^{\dagger}_{-\mu}(0;n)      \label{QP}
\end{equation}
For the  chiral condensate $<\bar\psi\psi>$ we then have
\begin{equation}
<\bar\psi\psi> =
{\mathop{\rm Tr}} G_\mu(0;0) =
{\mathop{\rm Tr}} G_{-\mu}^\star(0;0)     \label{CHI}
\end{equation}
where the second equality follows from eq. \ref{QP}.
Note that eq. \ref{CHI} implies that $<\bar\psi\psi>$ is real only in the full
ensemble average, when time reversal symmetry must
be realized.

In our particular simulation,
we used a noisy estimator
for $<\bar\psi\psi>$. So, in our case
eq. \ref{CHI} must hold only when the average over the
noise is taken. We thus lose this convergence test on
isolated configurations, but we  can  check a posteriori
the statistical quality of our sample by verifying (\ref{CHI})
for the ensemble.

Similar remarks hold for the charge operator
$<J_0>$, obtained by differentiating
the action with respect to the chemical potential. As discussed in
\cite{CHEM1} $<J_0>$ is the expectation value of the number of paths
in the t direction.

In addition to independent calculations for positive and negative
chemical potential for each configuration, we calculated all observables
for the three different $Z_3$ (antiperiodic) boundary conditions defined
by
\begin{equation}
\psi(t + N_t) = (-1)e^{i(2\pi k/3)}\psi(t); \hskip 0.5truecm k=(0,1,2)
\label{BOU}
\end{equation}
to enforce some of the constraints of confinement configuration by
configuration.

The spectrum computation is more delicate: in the meson sector
we have to compute
\begin{equation}
C^i_{q\bar q }(T) = \sum TrG_\mu(0;n)\Gamma_i G_\mu(n;0)(-1)^n
\end{equation}
where $\Gamma_i$ stands for the generic combination of gamma matrices
associated with each meson.
Inserting (\ref{QP})
we see that $C^i(t)$ should be computed according to
\begin{equation}
C^i_{q\bar q} (T) = \sum TrG_\mu(0;n)\Gamma_i G_{-\mu}^\dagger(0;n)
\end{equation}
(unless we want to compute the fermion propagator with a source at all points
of the lattice), and this requires the inversion of the Dirac operator
with opposite values of the chemical potential, representing the contributions
of quark and antiquark, respectively.

The same property (\ref{QP}) together with shift invariance
in $t$ gives the following symmetry for the propagators
$C_i$ which must hold for ensemble averages:
\begin{equation}
   C^i_{q\bar q} (T)  = C^i_{ q \bar q} (N_t - T) \label{MES}
\end{equation}
In the ensemble average we thus recover (at least in the
confined phase, where both of the above mentioned symmetries hold)
the usual parametrization
for the meson propagators.
The standard sum rule (Ward identity) holds configuration by configuration
in the modified form
\begin{equation}
<\bar\psi\psi>_\mu = m_q \sum_T C^\pi_{q\bar q}(T) \label{W1}
\end{equation}

\begin{equation}
<\bar\psi\psi>_{-\mu} = m_q \sum_T C^\pi_{\bar q q}(T) \label{W2}
\end{equation}
which gives again eq. \ref{CHI}
\begin{equation}
<\bar\psi\psi>_{\mu} = < \bar\psi\psi>_{-\mu}^\star
\end{equation}

For the nucleon things are different: the only exact relationship
in the ensemble average is:
\begin{equation}
C^N_{qqq}(t) = (-1)^T C^N_{\bar q\bar q \bar q} (N_t - T) \label{NUC}
\end{equation}
and no simple relationships exist between $C_{qqq}(T)$ and
$C_{qqq}(N_t-T)$.
In other words
for the baryon the usual parametrization is modified due to the
different behaviour in backward and forward propagation induced by
finite $\mu$: the ``minimal'' baryon propagator at finite
density contains at least
two positive parity excitations. It is clear that finite size effects are,
also from this point of view, especially severe.

In all our spectrum measurements we made use of a wall source \cite{WALL},
after the appropriate gauge fixing. In this way our propagators
reach their asymptotic regime faster, but we pay the price of
an amplification of the non-positivity effects connected with
finite density.

Also, for the spectrum computation we calculated all observables
for the three different $Z_3$ (antiperiodic) boundary conditions defined
in eq. \ref{BOU}. Since masses are non-linear observables, we may expect
that the results obtained by averaging the masses obtained
on subsamples corresponding to fixed boundary condition are different from
those obtained after averaging the propagators on all the three boundaries,
for our {\it finite} ensemble.
Such behaviour, if present,  would provide evidence of winding loops, which
have yet to cancel.

\subsection{Numerical analysis}

{}~~~~The theory at finite density was simulated on a $16^3\times 32$ lattice,
at bare quark mass $m_q = .02$ and $\beta =  6.0$.
For these parameters, the mass of the baryon at zero
density is $.77$ and the mass of the pion is $.34$. The
region of the chemical potential we have successfully explored ranges from
zero to $m_\pi/2 = .17$. We have also made some exploratory runs
at $\mu = .2$.

The results we discuss below result from 30 configurations
with the first boundary condition, 19 with the second , and 19
with the third analyzed
at $\mu = .0$, (30+19+19)$\times$ 2 at
$\mu = .1$, (30+30+30) $\times$ 2 at $\mu = .15$,
and (44+44+44)$\times$ 2 at $\mu = .17$, (26+26+26)$\times 2$ at $\mu = .2$
(recall that at $\mu \ne 0.$ we solve the Dirac equation for positive
and negative $\mu$).

Our configurations were generated by an admixture of Metropolis
and overrelaxed algorithms. We analyzed them every 10000 sweeps,
after initially discarding 12000 sweeps for thermalization.

We begin by discussing the behaviour of the chiral condensate and
number density. A few comments are
in order. First, the fluctuations
increase strongly with $\mu$. Second, the different boundaries
give rise to slightly different results even at $\mu = 0$.
Finally by increasing $\mu$  we observe several configurations in
which the results obtained with opposite $\mu$ values are completely
different. Clearly, a finite chemical potential amplifies the inhomogeneities
of the single configurations (note that such differences would vanish were we
to average over our noisy sources).

We have verified that the results for the chiral condensate and
number density obtained with $\pm \mu$, and with the
three different boundaries, are mutually consistent. We thus average
over them configuration by configuration,
and we show in Figs. \ref{fig1} and \ref{fig2} the resulting histograms.
Their main characteristic (which is common to  all the $\mu$ values)
is the absence of a two peak signal, which
would suggest a phase transition.
The results change smoothly from $\mu = 0$ to  $\mu = .17$, while at
$\mu = .2$ the results are much noisier.
 The regular structure
of the distributions gives us confidence that in the entire range
of $\mu$ studied here the system is in the chirally broken phase
(note, however, that most of the chiral condensate comes
from the explicit symmetry breaking mass term : at zero $\mu$,
$<\bar \psi \psi> = .13$ and only .03 is due to spontaneous symmetry
breaking \cite{SKIM}). From Figs. \ref{fig1} and \ref{fig2} we can also
appreciate the increasing width of the distributions with $\mu$,
and the presence of scattered events.
As a further consistency check, we also
performed partial analyses discarding those values which deviated
most strongly from the average. Nonetheless, we consistently found
compatible results. So, the situation is well under control
from the statistical point of view, and our data for
the chiral condensate and number density do not show any sign of a phase
transition.

The spectroscopic analysis posed more specific problems. As
stated above, we are dealing with non-positive definite operators.
Violation of positivity is also possible because of the wall source
we are using. These effects are so significant
that they even produced
a few pion propagators which are negative at zero distance!
Another dramatic feature in some of our propagators are huge fluctuations:
when that occurs, the shape of the propagator is altered as well.
This contrasts
with the situation at zero chemical potential, where amplitudes may
be fluctuating, even strongly, but the hyperbolic cosine behaviour
is preserved,  even, for instance,  in the `exceptional' configurations
observed with Wilson fermions near $\kappa_c$.
To be more specific, we show in Fig. \ref{fig3} a(b) the collection of the
pion (baryon)
propagators at $\mu = .15$, where the problem was observed first
(note that the exceptional propagator even has the `wrong' symmetry!).
In Fig. \ref{fig4}a(b) we show the same data at $\mu = .17$.
The change in behaviour while increasing $\mu$ is
dramatic: however, the expected hyperbolic
cosine pattern is still visible, and the
average propagators do not show qualitative pathologies.

What is the origin of these exceptional configurations?
What will ultimately occur in the limit of large statistics?
The most natural explanation is the occurrence of zero modes,
 and the
question to be answered concerns their physical significance.
As discussed in Sec.II, isolated zero modes
in quark propagators calculated on individual configurations can still
be compatible
with confinement and chiral symmetry breaking.

Here we want to suggest that (1.) the
origin of these exceptional configurations
may  be completely trivial, simply related
to statistical fluctuations, as anticipated in the Introduction;
and, (2.) to provide arguments which support their suppression
in the continuum, infinite volume limit.

To make our point clear, it is useful to characterize the
behaviour of a configuration by the effective masses, both
for the pion and baryon.
For the effective mass analysis we have extracted the direct
channel according to \cite{DIR}
\begin{equation}
G_{direct}(2t) = 2G(2t)+ G(2t+1) + G(2t-1)
\end{equation}
(We have systematically checked that the results of global fits give
compatible, although less accurate, results.)
 Also, for the baryon we
took into account the modified parametrization discussed above simply
by analyzing half of the lattice, which is justified by the fast
decay of the baryon correlators.

To begin,  we show in Fig.\ref{fig5}
the results from the effective mass
analysis at $\mu=0$ for half the  pion mass and for one third
the baryon mass performed on individual
configurations.
{}From Fig.\ref{fig5} we can see an overlap between half the pion mass, and one
third of the baryon mass, which is better demonstrated by the relative
histogram, Fig. \ref{fig6}.

Let us now consider the behaviour at $\mu =.17$, first for the pion mass
(we are using now only those propagators which give positive numbers for
the effective masses).
In Fig. \ref{fig7} we compare the distribution of (half the) pion mass at these
two
$\mu$ values: the distribution spreads out while increasing $\mu$,
an effect already observed
in the measurements of the chiral condensate and the number density.
Of course on the left the distribution  is bounded by zero, which
results in an asymmetric shape.

Analogously, we show in Fig.\ref{fig8}
the results for the baryon: in this case, in addition to the spreading,
we observe also a shift in its central value.
(We will see later on that the shift in one third of the
baryon screening mass is  $\mu$. Here we also plot  the distribution
shifted by $\mu$ which, modulo the spreading, coincides
with the one at $\mu = 0$.) Again, the left wing of the distribution
is  ``missing''.

In both cases (pion and baryon) we may associate the pathological
configurations
with the ones which should populate  the left part of the distribution.
The following scenario is certainly possible:
the pion mass distribution has
 half-width $\Delta = \Delta(m_q, L, \mu, \beta)$.
The pion mass (i.e. the central value of the distribution) does not
change in the confined phase. However,
the first zero modes
show up when $m_\pi - \Delta(m_q, L, \mu, \beta) = 0$,
around half the mass of the pion in this simulation. Since
$\lim_{L \to \infty, \beta \to \infty} \Delta = 0$
this  pathology is a  lattice artifact, and the quenched
theory should
make perfect sense in the continuum, infinite volume  limit.
An analogous argument can be made for
the baryon.

This mechanism, which is simply derived from the natural fluctuations
at finite size and spacing, is enough to account for the apparent early
onset of the chiral transition  reported in the past.
We cannot of course exclude that other more fundamental pathologies
affect  the theory at finite density. Only simulations on larger
lattices, and possibly closer to the continuum, can
definitively settle the issue.

We now turn to the conventional effective mass analysis.
Again, the results obtained for the three different boundary
conditions were fully consistent. We averaged over them.
At $\mu = .15$ and $.17$ we had to eliminate
the exceptional configurations (only one, actually, at $\mu = .15$)
in order  to obtain rather clean results for the effective masses.
We stress however that the results of global fits performed
on the full sample, although
very noisy, are compatible with those obtained by the
effective mass analysis on a selected subsample.
The results for the effective masses are shown in Fig. \ref{fig9} for the pion,
and \ref{fig10} for the baryon.

\subsection{Results.}

In table \ref {table1}
we report the results for the chiral condensate and the
number density. The data was averaged
over the three different boundaries, and over $\pm \mu$.
We quote also the imaginary parts, which  are consistent
with zero, as they should be. Fig. \ref {fig11} shows the corresponding plots.

Table \ref {table2} shows the results for the pion and baryon masses.
As discussed above,
at this stage in our ongoing project, we quote our results for the effective
mass analysis at $\mu = .15$ and $.17$, with the caveat that
they have been obtained on a subsample.
We can justify this procedure in part by noting that the results from the fits
on the full sample are fully consistent, with enlarged statistical errors,
with the ones we quote. The screening masses are plotted as a function of $\mu$
in Fig. \ref{fig12}.

The results can be summarized as following:
\begin{eqnarray}
J_0(\mu) & = & J_0(0) \nonumber \\
<\bar\psi\psi>(\mu) &  =  &<\bar\psi\psi>(0) \nonumber \\
m_{\pi} (\mu) & = & m_{\pi}(0) \nonumber \\
m_N (\mu) & = & m_N (0) - 3 \mu
\end{eqnarray}
This trend, if maintained, would give $\mu_c = m_N/3$.

Again, recall that the term $3 \mu$ in the baryon screening mass is
expected of simple quark models of nuclear matter.
They predict that the nucleon screening mass will decrease linearly with
increasing chemical potential $\mu$ and a chiral symmetry restoration
transition
will occur at $\mu = m_N/3$.

\section{Discussion and Prospects for Future Work}

{}~~~~In summary, we believe that the criticism and pathologies of quenched
QCD pointed out in the past can be interpreted in terms of
the fluctuations expected for $\mu
\put(0,-4){\shortstack{$>$\\$\sim$}} \hskip 0.4 truecm
 m_\pi/2$  as discussed in Sec. 2 above. It need not be true that quenched QCD
is unreliable at nonzero $\mu$. We believe, in fact, that the difficulties
in simulating quenched QCD at nonzero $\mu$ will be equally severe
in the full theory, but both classes of simulations will be ultimately
successful. The constraints of confinement are, we believe, absolutely
essential to obtain physical results from simulations at nonzero
chemical potential and, as we have argued above, the traditional simulation
scheme for lattice QCD is not well suited for this purpose.

We are now preparing a new set of simulations.
Our past measurements made use of a wall source for spectroscopy,
and of a noisy estimator for $\langle\bar{\psi}\psi\rangle$.
We are now testing a ``noisy'' wall source. Such a source is obtained
from our simple wall source by performing a random gauge transformation. This
source gives us
a stochastic estimator of the hadron propagators for a point source,
averaged over all points on the source time-slice. A point source
gives, in general, propagators which are more poorly behaved
than those produced by a wall source. However,
averaging over a large enough ensemble, a noisy source has the advantage of
also
averaging over all points on the source time-slice, increasing our effective
ensemble size by a factor equal to the number of {\it independent} point
sources on the time-slice. This approach should
increase the effective ensemble size and reduce the variance by enforcing
confinement. We believe such an effect is the reason why our stochastic
estimator for $\langle\bar{\psi}\psi\rangle$ is much better behaved when we
enter the region
$\mu
\put(0,-4){\shortstack{$>$\\$\sim$}} \hskip 0.4 truecm
 m_\pi/2$  than the hadron propagators
obtained from a simple wall source \cite{WALL}.

    We are also planning a simulation on a larger, $64 \times 16^3$,
lattice. Larger lattices should help control all the possible pathologies
discussed in Sec.II : the constraints of confinement are clearer
on larger lattices, variances in effective masses are diminished and
violations of symmetries are suppressed. Since we will also be using
the better measurement techniques discussed above, we are hopeful
that we will obtain more decisive simulation results for $\mu$
between $m_\pi/2$ and $m_N/3$.
In addition, the increase of $N_t$ further decreases the temperature ($1/N_t$)
of the lattice, which also helps suppress pathologies.
\vskip 2truecm
We would like to thank Ian Barbour and Eduardo Mendel for
interesting conversations.

This work was supported in
part by NSF via grant NSF-PHY92-00148 and
by DOE contract W-31-109-ENG-38. Simulations were done using the CRAY C-90
at NERSC.

\begin{table}
\begin {tabular} {l l l}
$\mu$ &  ${\mathop{\rm Re}}
<\bar{\psi}\psi>$ & ${\mathop{\rm Im}}<\bar\psi\psi>$   \\
 \hline
 0.000&  0.13769 (60)& -0.0006(10)\\
 0.100&  0.13750 (70)& -0.0021(11) \\
 0.150&  0.1362 (18)&  -0.0001(13)\\
 0.170&  0.1359 (19)&  -0.0051(21)\\
 0.200&  0.121   (4)&  -0.0027(65)\\
\hline
 $\mu$&  ${\mathop{\rm Re}}<J_0> $   &  ${\mathop{\rm Im}}< J_0> $   \\
 \hline
 0.000& -0.00047(98)& -0.00085(64)  \\
 0.100&  0.00033(83)& -0.00010(76) \\
 0.150&  0.00069(145)& -0.0022(12) \\
 0.170&  0.00071(162)& 0.0007(17) \\
 0.200&  0.004(5) &  0.0009(7) \\
 \end{tabular}
\caption{Results for the chiral condensate and the number density
as a function of $\mu$}
\label{table1}
\end{table}

\begin{table}
 \begin {tabular} {l l l }
 $\mu$ &  $m_\pi$ & $m_N$  \\
 \hline
 0.000&  0.3396(36)&  0.741(15)\\
 0.100&  0.3374(75)&  0.442(14)\\
 0.150&  0.3182(52)&  0.292(22)\\
 0.170&  0.313(15)&  0.235(24)\\
 \end{tabular}
\caption{Results for the pion and the baryon screening mass
as a function of $\mu$}
\label{table2}
\end{table}

\begin{figure}
\caption{Frequency plot for the chiral condensate
at $\mu = 0$ (a) and $\mu = .17$ (b).
 For each configuration
we averaged over the three boundaries, and the opposite $\mu$ values.}
\label {fig1}
\end{figure}

\begin{figure}
\caption{Frequency plot for the number density
$<J_0>$ at $\mu = 0 (a)$ and $\mu = .17$ (b).
For each configuration we averaged
over the three boundaries, and the opposite $\mu$ values.}
\label {fig2}
\end{figure}

\begin{figure}
\caption{ Pion(a) and baryon(b) propagators obtained with the
first boundary condition at $\mu = .15$}
\label{fig3}
\end{figure}

\begin{figure}
\caption{Same as Fig. 3 for $\mu = .17$ }
\label{fig4}
\end{figure}
\noindent

\begin{figure}
\caption{Effective masses computed configuration by configuration
for the $m_\pi/2$ (circles) and $m_B/3$ (squares) at $\mu = 0$.}
\label{fig5}
\end{figure}

\begin{figure}
\caption{Histograms accompanying Fig. 5. Dash is for $m_\pi/2$
and solid is for $m_N/3$.}
\label{fig6}
\end{figure}

\begin{figure}
\caption{Histograms  for $m_\pi/2$ at $\mu = .17$ (solid). The histogram
at $\mu = 0$ is shown for comparison (dash).}
\label{fig7}
\end{figure}

\begin{figure}
\caption{Histograms  for $m_N/3$ at $\mu = .17$ (solid). The
same, shifted by $\mu$ (dot).  The histogram
at $\mu = 0$ is shown for comparison (dash).}
\label{fig8}
\end{figure}

\begin{figure}
\caption{Effective masses for the pion as a function of time, for
$\mu = (0, 0.1, 0.15, 0.17)$, (crosses, diamonds, squares, circles).}
\label{fig9}
\end{figure}

\begin{figure}
\caption{Effective masses for the nucleon as a function of time, for
$\mu = (0, 0.1, 0.15, 0.17)$, (crosses, diamonds, squares, circles).}
\label{fig10}
\end{figure}

\begin{figure}
\caption{$<J_0>$ (a)  and $<\bar\psi\psi>$ (b) as a function of $\mu$.}
\label{fig11}
\end{figure}

\begin{figure}
\caption {Pion (crosses) and baryon
(circle) masses as a function of $\mu$. The straight line
is $y = m_B(0) - 3\mu$.}
\label{fig12}
\end{figure}

\end{document}